\documentclass[11pt,a4paper]{article}
\usepackage[utf8]{inputenc}
\usepackage[T1]{fontenc}
\usepackage{amsmath}
\usepackage{amsfonts}
\usepackage{graphicx}
\usepackage{caption}
\usepackage{booktabs}
\usepackage{multirow}
\usepackage{siunitx}
\usepackage{amssymb}
\usepackage{color}
\usepackage{colortbl}
\usepackage{siunitx}
\usepackage{booktabs}
\usepackage{longtable}
\usepackage[left=2.50cm, right=2.50cm, top=2.50cm, bottom=2.50cm]{geometry}
\usepackage[onehalfspacing]{setspace}
\usepackage{placeins}
\usepackage{authblk}
\usepackage{lipsum}
\usepackage{dsfont}
\usepackage{apacite} 				
\usepackage{natbib}	
\usepackage{rotating}

\usepackage[title,titletoc,toc,page]{appendix}
\appendixpageoff
\appendixtocoff

\title{Small area estimation with generalized random forests: Estimating poverty rates in Mexico}
\author[*]{Nicolas Frink}
\author[*]{Timo Schmid}
\affil[*]{\small{Institute of Statistics, Otto-Friedrich-Universit\"{a}t Bamberg, Bamberg, Germany}}
\date{}

\begin{document}

\maketitle
\onehalfspacing
\normalsize
\setlength{\parindent}{12pt}

\begin{abstract}
Identifying and addressing poverty is challenging in administrative units with limited information on income distribution and well-being. To overcome this obstacle, small area estimation methods have been developed to provide reliable and efficient estimators at disaggregated levels, enabling informed decision-making by policymakers despite the data scarcity. From a theoretical perspective, we propose a robust and flexible approach for estimating poverty indicators based on binary response variables within the small area estimation context: the generalized mixed effects random forest. Our method employs machine learning techniques to identify predictive, non-linear relationships from data, while also modeling hierarchical structures. Mean squared error estimation is explored using a parametric bootstrap. From an applied perspective, we examine the impact of information loss due to converting continuous variables into binary variables on the performance of small area estimation methods. We evaluate the proposed point and uncertainty estimates in both model- and design-based simulations. Finally, we apply our method to a case study revealing spatial patterns of poverty in the Mexican state of Tlaxcala.
\end{abstract}


\section{Introduction}\label{sec:1}
Poverty is a primary feature of development challenges, and reducing poverty is the first sustainable development goal (SDG) of the United Nations and one of the World Bank twin goals \citep{UN2015,WB2015}. However, the COVID-19 pandemic has dealt a severe blow to poverty reduction efforts, with poverty rates increasing for the first time in two decades \citep{WB20}. This effect is particularly pronounced in regions with already high poverty rates, such as Latin America, exacerbating the impoverishment of vulnerable populations. As a result, the World Bank's goal of ending poverty by 2030 seems increasingly unachievable without significant targeted policy actions \citep{WB2015}. To effectively combat poverty, governments must be able to identify areas with the greatest need for intervention, channeling resources, skills, and innovation to these regions. However, many areas lack information on income distribution and well-being, making it difficult to identify and address poverty. This lack of information is often due to limited funding for population surveys, particularly for smaller spatially disaggregated areas such as states, districts, or municipalities \citep{Tzavidis2018,ADB2021}. Small area estimation (SAE) methods were developed to provide reliable and efficient estimators on disaggregated levels that allow policymakers to make informed decisions despite the lack of available information \citep{Molina2015}. The head count ratio (HCR) is an established economic indicator representing the percentage of households living below a certain poverty line within a given region. The HCR is based on a binary variable \citep{Peragine2021}, taking a value of $1$ if the household income is below the poverty line $t$ and $0$ otherwise \citep{Foster1984}. The HCR for an area $i$ is defined as:
$$\text{HCR}_i = \frac{1}{N_i} \sum_{j=1}^{N_i} \mathds{1}(z_{ij} \leq t),$$
where $\mathds{1}(\cdot)$ denotes the indicator function, $z_{ij}$ represents the income of household $j$ and $N_i$ is the population size of area $i$. Previous literature on SAE has estimated this binary-based poverty indicator using models that assume continuous input variables, such as household income. Limited access to continuous unit-level household income data for reasons of data security poses a challenge for SAE practitioners. However, using input variables that merely indicate whether a household is categorized as poor can help mitigate this issue.

A variety of methods for SAE with binary outcomes have been explored. For instance, \citet{Malec1997} and \citet{Nandram1999} propose a hierarchical Bayes (HB) approach. Alternatively, empirical Bayes (EB) methods, as described by \citet{MacGibbon1989} and \citet{Farrell1997}, offer another approach. From a frequentist perspective, the empirical best predictor (EBP) suggested by \citet{Jiang2001} and \citet{Jiang2003} represents a viable option. Moreover, \citet{Chambers2016} propose M-quantile modeling as a robust alternative for binary outcomes, offering protection against model misspecification and outliers. However, each of these methods assumes a linear relationship within the systematic component of the underlying model.

Hence, this paper introduces the generalized mixed effects random forest (GMERF) within the methodological tradition of SAE as a novel, flexible, data-driven, and semi-parametric approach for estimating poverty rates based on binary variables. We incorporate the random forest \citep{Breiman2001} within the mixed models framework because it exhibits excellent predictive performance, captures relationships directly from the data, and can incorporate higher-order interactions between covariates without requiring explicit model assumptions \citep{Hastie_etal2009, Biau2016}.

Machine learning methods, in general, have already been applied in SAE \citep{Bilton2017, Bilton2020}. Nonetheless, they exhibit limitations, such as disregarding the correlation among subpopulations and the incapacity to handle intricate covariance structures. Recently, there has been a growing interest in expanding the application of one particular machine learning technique, specifically the tree-based method, by incorporating it into mixed effects models. This integration allows for the surpassing of the inherent limitations associated with these models. For instance, \citet{Krennmair2023} and \citet{Krennmair2022} introduce the concept of mixed effect random forests (MERF) within the methodological framework of SAE. They assume a continuous response variable and substitute the linear combination of covariates in the fixed effects section of a linear mixed model (LMM) with a regression forest. This approach is appropriate for Gaussian response variables but not suitable for classification problems. For binary response variables and individual predictions, different methods exist. For example, \citet{Hajjem2017} propose a generalized mixed effects regression tree (GMERT), while \citet{Fontana2021} present the generalized mixed effects tree (GMET). GMET utilizes the tree leaves as indicator variables instead of the tree predictions used in GMERT. Although existing methods extend the use of simple trees for modeling nested data with non-Gaussian response variables, they do not incorporate tree ensembles. Therefore, \citet{Pellagatti2021} extend the GMET approach \citep{Fontana2021} and use random forests instead of standard trees in the fixed effects section of the mixed effects model. Although the aforementioned methods combine machine learning and mixed models for estimating binary variables, they mainly focus on individual predictions.

The major methodological contribution of our paper is the extension of the GMERT approach \citep{Hajjem2017} to a GMERF and incorporating it into the SAE framework for estimating area-level proportions, for instance poverty rates. Furthermore, we propose a parametric mean squared error (MSE)-bootstrap scheme to asses the uncertainty associated with area-level estimates. In contrast to the predominantly regression-based 'traditional' SAE methods for binary target variables \citep{Jiang2001, Jiang2003}, the GMERF offers protection against model misspecification and captures non-linear relationships from data. This highlights a major drawback of the classical SAE methods, as they rely on model assumptions that barely match real findings when considering data on social and economic inequality. This discrepancy between assumptions and reality can introduce bias in parameter estimates, making MSE estimates unreliable \citep{JiangRao2020}.

In addition to our methodological contribution, we also aim to discuss the impact of information loss caused by converting continuous input variables into binary variables on the performance of estimation methods. To achieve this, we evaluate the performance of our GMERF against established SAE methods that use continuous household income as an input variable. These methods include the EBP with data-driven transformation \citep{Rojas2020} and the MERF for non-linear indicators \citep{Krennmair2022}.

The paper is structured as follows: Section 2 introduces GMERFs as a method that integrates random forests and hierarchical modeling to account for dependencies between unit-level observations. In Section 2.2, we elucidate the construction of area-level proportion estimates and deliberate upon the specific data scenarios in which our method works particularly well. To address the issue of accurately estimating the MSE of the area-level estimates, Section 3 proposes a parametric bootstrap scheme. In Section 4, we evaluate and compare the performance of the proposed GMERF method for point and MSE estimates with established SAE methods for binary target variables. In Section 5, we apply our proposed GMERF method to estimate area-level poverty rates and corresponding uncertainty estimates for the Mexican state of Tlaxcala. We present the modeling and robustness properties of our proposed methods and compare its performance with the EBP using data-driven transformation and MERF methods, considering variations (binary vs. continuous) in the available information about the target variable. Section 6 presents a design-based simulation to evaluate the quality of the results obtained in Section 5, providing a comprehensive demonstration of the properties and advantages of GMERFs in the context of SAE. Finally, Section 7 concludes our study and motivates further research in the field of SAE.

\section{Theory and method}\label{sec:2}
In this section, we propose a novel, flexible, and data-driven approach that extends previous methods. Our method utilizes a random forest algorithm to estimate area-level proportions in the presence of unit-level survey and census data.

\subsection{Generalized mixed effects random forests}\label{sec:2.2}
Consider a finite population $U$ that is divided into $D$ areas. $U_i$ denotes the population of the $i$-th area with $i=1,\dots, D$ and $N_i$ denotes the population size of the $i$-th area. The overall population size is given by $N = \sum_{i=1}^{D} N_i$. The number of observations sampled in area $i$ is denoted by $n_i$ and the total sample size is given by $n=\sum_{i=1}^{D}n_i$. Within each sampled area, $j$ individual observations are obtained, where $j$ ranges from 1 to $n_i$. The binary response variable for area $i$ is represented by a vector of individual observations $y_i = [y_{i1},\dots, y_{in_i}]'$ with dimension $n_i \times 1$. Let $X_i = [x_{i1},\dots, x_{in_i}]'$ and $Z_i = [z_{i1},\dots, z_{in_i}]'$ represent the $n_i \times p$ matrix of covariates and the $n_i \times q$ matrix of domain-specific random effect specifiers, respectively. Here, $p$ denotes the number of covariates and $q$ denotes the dimension of the random effects. The $q\times 1$ vector of random effects for area $i$ is denoted by $\nu_i=[\nu_{i1},\dots, \nu_{iq}]'$ which we assume normally distributed with the variance-covariance matrix $H_i$ for random effects of each domain $i$. The notation used for all areas is as follows: $y = \text{col}_{1\leq i \leq D}(y_i) = (y'_i,\dots,y'_D)'$, $X = \text{col}_{1\leq i \leq D}(X_i)$, $Z = \text{diag}_{1\leq i \leq D}(Z_i)$, $H = \text{diag}_{1\leq i \leq D}(H_i)$ and $\nu = \text{col}_{1\leq i \leq D}(\nu_i)$.

The objective is to predict values for non-sampled observations using accessible supplementary covariates from census data across domains. To achieve this, we seek to establish a relationship $f()$ between the covariates and the target variable. The fixed part, denoted as $f()$, represents the random forest and expresses the conditional mean of the linear predictor $\eta$ given the covariates $X$. The random part, represented by $Z\nu$, accounts for the dependencies introduced by random effects. \(\mu\) denotes the expected value of the response variable, given the random component, and is computed using the cumulative distribution function of the logistic distribution. Following \citet{Pellagatti2021}, the GMERF model is described in equation (\ref{mod1}):
\begin{align}\label{mod1}
\eta & = f(X) + Z \nu, \\
\nu &\sim N(0,H), \nonumber \\
\mu & = \frac{exp(\eta)}{1+ exp(\eta)}, \nonumber\\
E(y|\nu) & = \mu. \nonumber
\end{align}
The model in equation (\ref{mod1}) can be simplified to a generalized linear mixed model (GLMM) by setting $f(X)=X\beta$, where $\beta = [\beta_1,\dots,\beta_p]'$ are the regression parameters. It is important to note that inference based on GLMMs poses computational challenges due to the involvement of high-dimensional integrals in the likelihood, which cannot be evaluated analytically \citep{Stroup2012}. To overcome this issue, a popular approximation method is the penalized quasi-likelihood (PQL) approach. The PQL approach constructs a linear approximation of the distribution for non-normal response variables and assumes that the linearized dependent variable is approximately normally distributed. This gives the integration a closed form, allowing the use of maximum likelihood estimation in the algorithm \citep{Breslow1993, Stroup2012}. By employing the PQL approach to estimate a GLMM, we can derive a weighted MERF pseudo-model that uses a linearized target variable $y_L$. In order to linearize the binary response variable $y$, we apply a first-order Taylor series expansion, resulting in the transformed variable:
\begin{align}
	y_{L}=\text{ln}\Biggl(\frac{\mu}{1-\mu}\Biggr) + (y-\mu)\Biggl(\frac{1}{\mu(1-\mu)}\Biggr). \nonumber
\end{align}
The weighted MERF pseudo-model based on the MERF by \citet{Krennmair2023} is defined as follows:
$$y_{L} = f(X) + Z\nu+ \epsilon,$$
with $\epsilon \sim N(0, W^{-1})$, $\epsilon = \text{col}_{1\leq i \leq D}(\epsilon_i)$, $\epsilon_i = [\epsilon_{i1},\dots, \epsilon_{in_i}]'$ is the $n_i\times1$ vector of individual error terms and
$W = \text{diag}_{1\leq i \leq D}(W_i)$ are the weights. 
Note that in the special case of binary data, the variance-covariance matrix for the individual errors is equivalent to the inverse of the weights. These linearization weights are also used within the random forest algorithm for sampling training observations. Observations with larger weights will be selected with a higher probability in the bootstrap samples for the trees. For each area $i$, the following holds:
\begin{align*}
	y_{L_i}& = f(X_i) + Z_i\nu_i+ \epsilon_i\\
	\eta_i & = f(X_i) + Z_i \nu_i \nonumber \\
		\mu_i & = \frac{exp(\eta_i)}{1+ exp(\eta_i)}\nonumber\\
	E(y_i|\nu_i) & = \mu_i. \nonumber
\end{align*}
In the weighted MERF model, we make the assumption that $\nu_i$ and $\epsilon_i$ are independent normally distributed, and the between-area observations are also independent. The covariance matrix of the linearized variable $y_L$ can be expressed as $Cov(y_L) = V=\text{diag}_{1\leq i \leq D}(V_i)$, with $V_i = Z_iH_iZ_i' + W_i^{-1}$. Additionally, we assume that correlations in the weighted MERF pseudo-model arise solely from the between-domain variation, i.e., $W_i^{-1}$ is diagonal ($W_i = \text{diag}(w_{ij})$ with $w_{ij} = \mu_{ij}(1-\mu_{ij}))$.

The proposed GMERF-algorithm for fitting model (\ref{mod1}) is a doubly iterative process with micro iterations within macro iterations. In each macro iteration, the linearized response variable and weights are updated, while the micro iterations use an approach similar to the expectation-maximization (EM)-algorithm \citep{Moon1996} for parameter estimates. The updated linearized response variable and weights values serve as response variable and weights, respectively:
\begin{enumerate}
	\item Set $B=0$. Given initial estimates of the mean values $\hat{\mu}^{(B)}$ ($\hat{\mu}^{(B)}=0.75$ for $y=1$ and $\hat{\mu}^{(B)}=0.25$ for $y=0$), fit a weighted MERF pseudo model using the linearized response $y^{(B)}_{L}$ and the weights $W^{(B)}$.
	\begin{enumerate}
		\item Initialize $b=0$ and set random components $\hat{\nu}_{(0)}$ to zero.
		\item Set $b=b+1$. Update $y^*_{L(b)}$, $\hat{f}(X)_{(b)}$ and  $\hat{\nu}_{(b)}$:
		\begin{enumerate}
			\item $y^*_{L(b)} = y^{(B)}_{L} -Z\hat{\nu}_{(b-1)}$ (decorrelate the dependent variable).
			\item Estimate $\hat{f}()_{(b)}$ using a random forest with dependent variable $y^*_{L(b)}$, covariates $X$ and weights.
			\item Get the Out-of-Bag-predictions (OOB-predictions) from the random forest $\hat{f}(X)^{OOB}_{(b)}$.
			\item Fit a linear mixed model, with weights and restricted regression coefficient of 1 for $\hat{f}(X)^{OOB}_{(b)}$:
			\begin{align*}
				y^{(B)}_{L}=\hat{f}(X)^{OOB}_{(b)}+Z\hat{\nu}_{(b)} + \epsilon.
			\end{align*}
			\item Extract the variance components and estimate the random effects by:
			\begin{align*}
				& \hat{\nu}_{(b)}=\hat{H}_{(b)} Z'\hat{V}^{-1}_{(b)}(y^{(B)}_{L}-\hat{f}(X)^{OOB}_{(b)}).
			\end{align*}
		\end{enumerate}
		\item Repeat step (b) until convergence is reached in term of generalized log-likelihood (GLL) value:
		\begin{align*}
			GLL(f, v_i|y)=&\sum_{i=1}^{D}[(y_L-\hat{f}(X_i) - Z_i\hat{\nu}_{i})'\hat{W}_{i}\\&\times(y_L-\hat{f}(X_i) - Z_i\hat{\nu}_{i})+\hat{\nu}_{i}'\hat{H}^{-1}_{i}\hat{\nu}_{i} \\&+ \text{log}|\hat{H}_i| + \text{log}|\hat{W}^{-1}_{i}|].
		\end{align*}
	\end{enumerate}
	\item Set $B=B+1$: Update $\hat{\eta}$, $\hat{\mu}$, $\hat{y}_L$ and $w$:
	\begin{align*}
	\hat{\eta}^{(B)} &= \hat{f}(X)^{OOB} + Z\hat{\nu},\\
	\hat{\mu}^{(B)}&=\frac{exp(\hat{\eta}^{(B)})}{1+exp(\hat{\eta}^{(B)})},\nonumber \\
	y^{(B)}_{L} &= \text{ln}\Biggl(\frac{\hat{\mu}^{(B)}}{1-\hat{\mu}^{(B)}}\Biggr)+(y-\hat{\mu}^{(B)})\Biggl(\frac{1}{\hat{\mu}^{(B)}(1-\hat{\mu}^{(B)})}\Biggr),\nonumber \\
	w^{(B)} &= \hat{\mu}^{(B)}(1-\hat{\mu}^{(B)}),\nonumber
	\end{align*}
	where $\hat{f}(X)^{OOB}$ and $\hat{\nu}$ equal their estimated values at the micro-level convergence of the previous macro iteration.
	\item Repeat step 2 until $\hat{\eta}$ converges.
\end{enumerate}
For given $V$ and $f = X \beta$, the maximization of the GLL-criterion is equivalent to the solution of the mixed model equations \citep{Wu2006}. For given variance components, this leads to an estimator of the best linear unbiased predictor (BLUP) for the GMERF \citep{Gonzalez_etal2008}:
\begin{align}\label{mod3}
	\hat{\nu} = HZ'V^{-1}(y_L-\hat{f}(X)).
\end{align}

\subsection{Small area proportions}\label{sec:2.3}
Assuming the same simplifications proposed by \citet{Battese1988} throughout the paper, i.e., $q=1$, where $Z$ is a $n_i \times D$ design-matrix of area-intercept indicators, $\nu=[\nu_1, \dots, \nu_D]'$ is a $D \times 1$ vector of random effects, and the variance-covariance matrix of random effect simplifies to $H_i =\sigma^2_{\nu}$, the mean-estimator for each area $i$ based on available supplementary data sources can be calculated as follows:
\begin{align*}
	\bar{\hat{f}}(X_i) &= \frac{1}{N_i} \sum_{j=1}^{N_i}\hat{f}(X_i) = \frac{1}{N_i} \sum_{j=1}^{N_i}\hat{f}(x_{ij}).
\end{align*}
Since $\hat{\nu}_i$ is the BLUP, the proposed estimator for the area-level proportion is given by:
\begin{align}
	\hat{\eta}_i &= \bar{\hat{f}}(X_i) + Z_i \hat{\nu}_i, \nonumber\\
	\hat{\mu}_i &= \frac{exp(\hat{\eta}_i)}{1+exp(\hat{\eta}_i )}\text{ for $i = 1,\dots, D$.}
\end{align}
For non-sampled areas, the proposed estimator for the area-level proportion simplifies to the fixed component obtained from the random forest:
\begin{align*}\label{mod5}
	\hat{\eta}_i &= \bar{\hat{f}}(X_i), \\
	\hat{\mu}_i &= \frac{exp(\hat{\eta}_i)}{1+exp(\hat{\eta}_i )}\text{.}
\end{align*}
Unlike existing SAE approaches for binary target variables, our proposed estimator circumvents issues related to model selection. This advantage stems from the utilization of the random forest method, which implicitly incorporates optimized model selection, encompassing higher-order effects and non-linear interactions. Moreover, our estimator is adept at handling high-dimensional covariate data, even when the number of covariates surpasses the sample size \citep{Hastie_etal2009}. The predictive performance of our method is influenced by two crucial tuning parameters: the number of split-candidates at each node, which regulates the degree of decorrelation, and the number of trees.

In the subsequent discussion, we aim to explore the data situations in which our GMERF approach (for binary variables) may exhibit great performance and the potential to outperform the MERF (for continuous variables) when the interest is in estimating area-level proportions. This investigation will be carried out through application and design-based simulations in Section \ref{sec:5} and Section 6, focusing on comparing the performance of our proposed approach with established SAE methods that utilize continuous information, which inherently provides more information for the target variable. The PQL approach used in our algorithm works particularly well when the random effects are normally distributed \citep{McCulloch1997}. However, \citet{Krennmair2023} demonstrated the robustness of the MERF (whose properties we incorporate into our method) against misspecification of distributional assumptions in mixed models. Given that the GMERF relies on an approximation method, it is advisable to follow a rule of thumb similar to the central limit theorem when approximating a binary variable using the normal distribution. As suggested by \citet{Breslow2004}, a general guideline could be to ensure that the expected number of successes and failures exceeds $5$ in order to achieve a good approximation. Finally, we can draw upon an argument from the classical linear model (without random effects). In certain cases, a binary indicator is estimated using a linear probability model (LPM) that does not rely on a link function. This procedure is viable because the LPM often exhibits comparable performance to the traditional logit model, particularly when the probabilities fall within a specific range where they exhibit almost linear behavior with respect to the log-odds function. Generally, the LPM can be employed when the estimated probabilities range lies between $0.2$ and $0.8$ \citep{Long1997}. Hence, we assume that our approximation method exhibits satisfactory performance when the area-level probabilities predominantly fall within this range.

\section{Uncertainty estimation}\label{sec:3}
In the context of small area estimators, having an accuracy measure, typically expressed as MSE, is crucial. However, computing the analytical form of the MSE is not feasible even for relatively simple models like GLMMs. To address this, \citet{Gonzalez_etal2008} propose a rough approximation by linearizing the model and applying the Prasad-Rao approximation \citep{Prasad_Rao1990} for linear mixed models. Alternatively, bootstrap schemes provide a straightforward approach. In this paper, we propose a parametric bootstrap to estimate the MSE of the small area estimator introduced in equation (3). Our bootstrap scheme is built upon the method proposed by \citet{Gonzalez_etal2008} for small area estimators based on a logit mixed model. This approach focuses on capturing the dependence structure of the data and the uncertainty arising from the model estimation. The bootstrap procedure consists of the following steps:
\begin{enumerate}
	\item For $b = 1,\dots, B$:
	\begin{enumerate}
		\item Generate bootstrap random effects for the $D$ areas as $\nu_i^{(b)} \sim N(0, \hat{\sigma}_{\nu}^2)$ for $i=1,\dots, D$.
		\item Generate a bootstrap population of independent Bernoulli realizations made up of $D$ areas of sizes $N_i$, and with bootstrap Bernoulli realization $y_{ij}^{(b)}$ in area $i$ taking the value $1$ with probability:
		\begin{align*}
			\mu_{ij}^{(b)} = \frac{exp(\hat{f}(x_{ij}) + Z_i\nu_i^{(b)})}{1+ exp(\hat{f}(x_{ij}) + Z_i\nu_i^{(b)})}.
		\end{align*}
		\item Calculate the true bootstrap population area probabilities $\mu^{(b)}_{i}$ as  $\frac{1}{N_i}\sum_{j=1}^{N_i} y^{(b)}_{ij}$ for all $i=1,\dots,D$.
		\item For each bootstrap population draw a bootstrap sample with the same $n_i$ as the original sample. Use the bootstrap sample to obtain estimates $\hat{f}^{(b)}()$ and $\hat{\nu}^{(b)}()$.
		\item Calculate area-level probabilities by:
		\begin{align*}
			\hat{\mu}^{(b)}_i &=\frac{exp(\bar{\hat{f}}^{(b)}(X_i)+Z_i\hat{\nu}^{(b)}_i)}{1+exp(\bar{\hat{f}}^{(b)}(X_i)+Z_i\hat{\nu}^{(b)}_i)}.
		\end{align*}
	\end{enumerate}
	\item Using the $B$ bootstrap samples, the MSE estimator is obtained as follows:
	\begin{align*}
		\widehat{MSE}^{param}_i = \frac{1}{B}\sum_{b=1}^B(\hat{\mu}^{(b)}_i - \mu^{(b)}_{i})^2.
	\end{align*}
\end{enumerate}

\section{Model-based simulation study}\label{sec:4}
This section marks the initial phase of our empirical evaluation of the proposed method. To assess its performance, we employ a model-based simulation that compares point estimates for area-level proportions derived from the GMERF model in equation (\ref{mod1}) with those obtained from several competing models. Specifically, we investigate the performance of GMERFs in comparison to established SAE methods for binary outcomes, such as the conditional expectation predictor (CEP) and the M-quantile (MQ) approach \citep{Chambers2016}. By contrasting the performance of these linear competitors with our more flexible approach, which incorporates semi-parametric and non-linear modeling, we aim to showcase the advantages of our methodology. The primary objective is to demonstrate that our proposed method, in terms of point and uncertainty estimates, performs comparably well to traditional SAE methods while exhibiting comparative strengths in terms of capturing non-linear relations in the data.

The simulation setting is defined by a finite population $U$ of size $N=50000$, comprising $D=50$ disjoint areas $U_1, ..., U_D$ of equal size $N_i=1000$. To generate samples, we employ stratified random sampling, treating the $50$ small areas as strata. This results in a sample size of $n = \sum_{i=1}^D n_i = 687$. The sample sizes for each area range from $1$ to $28$ sampled units, with a median of $13$ and a mean of $14$. These sample sizes align with the area-level sample sizes observed in the application described in Section 5.

We contemplate four scenarios: \textit{Normal-Small}, \textit{Interaction-Small}, \textit{Normal-Large}, and \textit{Interaction-Large}. Each scenario is independently repeated $M = 500$ times. By comparing the estimates obtained from different models in these scenarios, we can assess their performance when faced with unknown non-linear interactions between covariates and varying levels of variation in the random effect. The \textit{Normal} scenarios serve as a reference for the CEP and MQ models. Since the model assumption of linearity is fulfilled, we aim to show that GMERFs perform comparably well to linear competitors in the reference scenario. The \textit{Interaction} scenarios differ from the \textit{Normal} ones as they involve a more complex model that incorporates quadratic terms and interactions on the linear predictor scale. These scenarios demonstrate the advantages of semi-parametric and non-linear modeling methods that protect against model-failure. Within each of these two scenarios, we explore two specifications for the random effects: small and large. These specifications account for different levels of magnitude in the between-group variability, thereby providing insights into the performance of the models under varying levels of variability. The indication of a small or large random effect is always to be put in relation to the given effect strength of the fixed effects. To illustrate this more precisely, we calculate the variance partition coefficient (VPC) \citep{Goldstein2002}. The VPC serves as a measure of intraclass correlation. In the context of binary outcomes, it indicates the explanatory variance share on the target variable that is to be assigned to the classification by the group variable. The VPC can be calculated as follows:
\begin{equation*}
	VPC = \frac{\sigma^2_{\nu}}{\sigma^2_{\nu} + \sigma^2_{latent}},
\end{equation*}
where $\sigma^2_{latent}$ denotes the residual variance that cannot be explained by either the fixed effects or the group-specific random intercepts. This latent variance can be estimated by the variance of the logistic distribution, which is $\sigma^2_{latent}= \frac{\pi^2}{3}$, with $\pi\approx3.14159$, as demonstrated in the fixed effects logistic regression model \citep{Browne2005, Fontana2021}. In this study, we use the terms small or large to describe scenarios with a VPC of $0.03$ or $0.23$, respectively. A detailed description of the data-generating process for each scenario is available in Table \ref{tab:MB1}.

\begin{table}[ht]
	\centering
	\captionsetup{justification=centering,margin=1.5cm}
   \caption{Model-based simulation scenarios}
	\resizebox{\textwidth}{!}{\begin{tabular}{rlccccccc}
			\toprule
			{Scenario} & {Linear predictor} & {Probability} & {$y$} & {$x1$} & {$x2$} & {$\nu$} \\ \midrule
				Normal-Small  & $ \eta = 0.5-0.8x_1-0.6x_2+\nu$& $\mu=exp(\eta)(1+exp(\eta))^{-1}$& $ Bin(\mu)$& $N(0,2^2)$ & $N(0,3^2)$  & $N(0,0.1)$\\
			Interaction-Small  & $ \eta = 1-x_1x_2-0.6x_1^2+\nu $ &$\mu=exp(\eta)(1+exp(\eta))^{-1}$&$ Bin(\mu)$ & $N(0,2^2)$ & $N(0,3^2)$  & $N(0,0.1)$  \\
			Normal-Large  & $ \eta = 0.5-0.1x_1-0.2x_2+\nu $ &$\mu=exp(\eta)(1+exp(\eta))^{-1}$&$ Bin(\mu)$ & $N(0,2^2)$ & $N(0,3^2)$  &  $N(0,1)$ \\
			Interaction-Large  & $ \eta = 1-x_1x_2-0.6x_1^2 +\nu$&$\mu=exp(\eta)(1+exp(\eta))^{-1}$& $ Bin(\mu)$ & $N(0,2^2)$ & $N(0,3^2)$   & $N(0,1)$  \\ \bottomrule
	\end{tabular}}
\label{tab:MB1}
\end{table}
We assess the quality of point estimates for area-level proportions using two metrics: relative bias (RB) and relative root mean squared error (RRMSE). To evaluate the accuracy of the MSE estimates, we consider the relative bias of the root mean squared error (RB-RMSE) and the relative root mean squared error of the RMSE:
\begin{align}
\nonumber	RB_i &= \frac{1}{M} \sum_{m=1}^{M} \left(\frac{\hat{\mu}^{(m)}_i - \mu^{(m)}_i}{\mu^{(m)}_i}\right)\\\nonumber
\textcolor{black}{RRMSE_i} &= \textcolor{black}{\frac{\sqrt{\frac{1}{M} \sum_{m=1}^{M} \left(\hat{\mu}^{(m)}_i - \mu^{(m)}_i\right)^2}}{\frac{1}{M}\sum_{m=1}^{M}\mu^{(m)}_i}}\\\nonumber
RB\text{-}RMSE_i &=\frac{\sqrt{\frac{1}{M} \sum_{m=1}^{M} MSE^{(m)}_{est_i}} - RMSE_{emp_i}}{RMSE_{emp_i}}\\\nonumber
RRMSE\text{-}RMSE_i &= \frac{\sqrt{\frac{1}{M} \sum_{m=1}^{M} \left(\sqrt{MSE^{(m)}_{est_i}} - RMSE_{emp_i}\right)^2}}{RMSE_{emp_i}},
\end{align}
where $\hat{\mu}^{(m)}_i$ represents the estimated proportion in area $i$ obtained from any of the methods mentioned above and $\mu^{(m)}_i$ denotes the true proportion for area $i$ in simulation round $m$. The estimation of $MSE_{est_i}^{(m)}$ is performed using the proposed bootstrap method described in Section \ref{sec:3}. Furthermore, $RMSE_{emp_i}$ is calculated as the empirical root mean squared error over $M$ replications, given by $\sqrt{\frac{1}{M} \sum_{m=1}^{M}(\hat{\mu}^{(m)}_i -\mu^{(m)}_i)^2}$.

To implement the model-based simulation, we use R \citep{R_language}. The CEP estimates are obtained using the \emph{lme4} package \citep{Bates_etal2015}, while the \emph{BinaryMQ} package \citep{Chambers2016} is used to compute the MQ estimates. For the proposed GMERF approach, we employ the \emph{ranger} \citep{Wright17} and \emph{lme4} \citep{Bates_etal2015} packages. To monitor the convergence of the algorithm, we use a precision of $1e^{-5}$ in the relative difference of the GLL-criterion and a precision of $0.01$ in the relative change of $\hat{\eta}$. 

\begin{figure}[ht]
	\centering
	\captionsetup{justification=centering,margin=1.5cm}
		\includegraphics[width=1\linewidth]{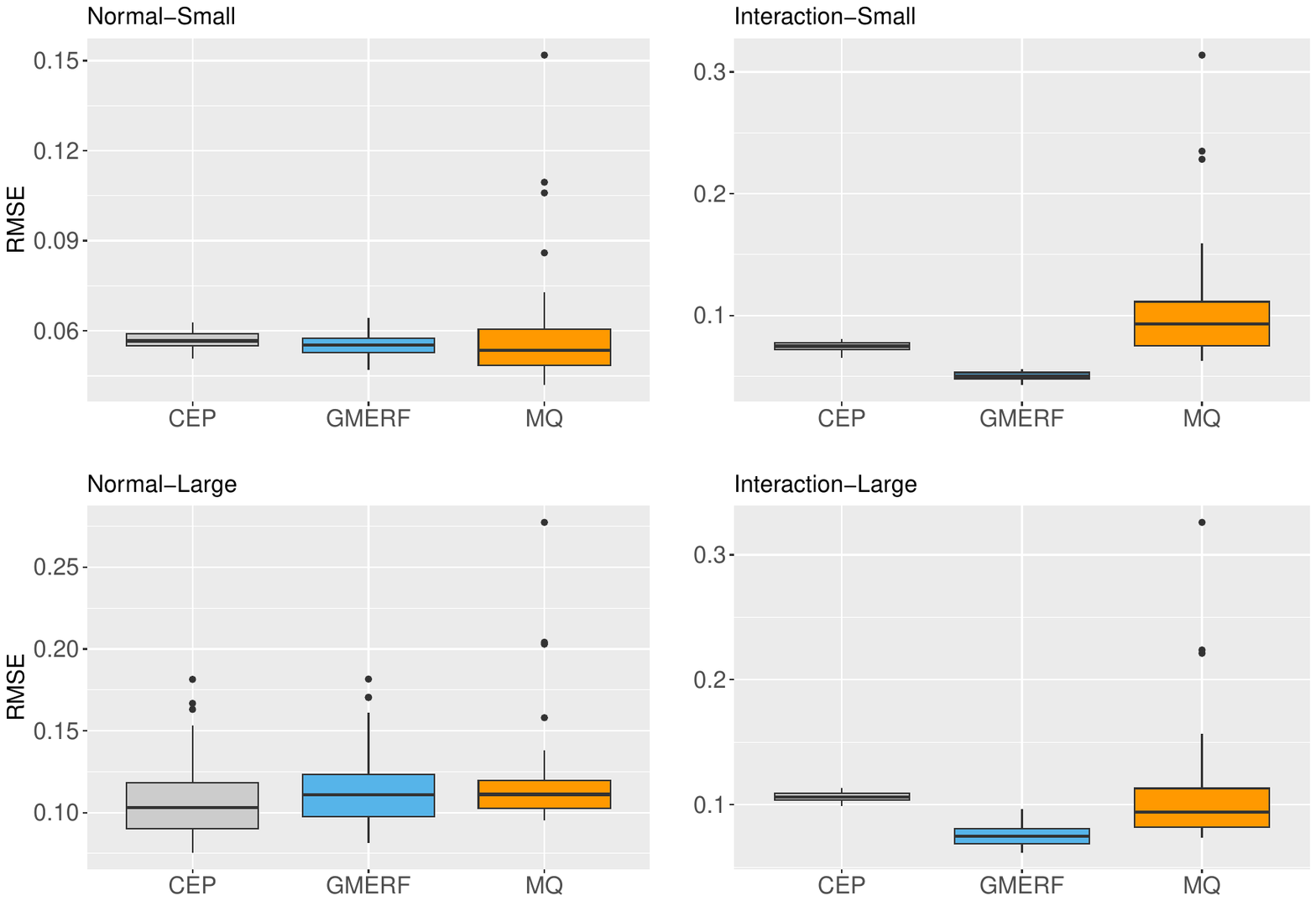}
	\caption{Empirical RMSE comparison of point estimates for area-level proportions under four scenarios}
	\label{fig:MBpoint}
\end{figure}

Figure \ref{fig:MBpoint} displays the empirical RMSE of each method across the four scenarios. As anticipated, in the \textit{Normal-Small} scenario, the GMERF method does not surpass the CEP and MQ estimators, but instead performs comparably. A similar pattern is observed in the \textit{Normal-Large} scenario, with all three estimators exhibiting a higher overall RMSE. The competitors conform to the fixed effects of the data-generating process and perform accordingly. For complex scenarios, namely \textit{Interaction-Small} and \textit{Interaction-Large}, point estimates from the proposed GMERF method outperform the competitors. In the \textit{Interaction-Small} scenario, the CEP has lower RMSE values than the MQ estimator. However, in the \textit{Interaction-Large} scenario, the MQ estimator outperforms the CEP in terms of lower RMSE. The flexible GMERF approach automatically identifies interactions and non-linear relationships, such as quadratic terms, leading to an advantage in terms of RMSE. Overall, the results in Figure \ref{fig:MBpoint} indicate that the GMERF performs comparably well in linear scenarios and outperforms traditional SAE-models in the presence of unknown non-linear relationships. Table \ref{tab:MBpoint} reports the corresponding values of RB and RRMSE for each discussed point estimate. The RB and RRMSE of the GMERF method exhibit a competitive low level across all scenarios. Specifically, in the \textit{Interaction-Large} scenario, a well-known observation regarding the statistical properties of random forests becomes apparent: the RB is slightly higher compared to the CEP, yet  this increased RB is compensated by a lower RRMSE for point estimates.

\begin{table}[!h]
\footnotesize
	\centering
	\captionsetup{justification=centering,margin=1.5cm}
	\caption{Mean and median of RB and RRMSE over areas for point estimates}
	\begin{tabular}{@{\extracolsep{5pt}} lrcccccccc}
			\\[-1.8ex]\hline
			\hline \\[-1.8ex]
			& &\multicolumn{2}{c}{\textit{Normal-Small}} &\multicolumn{2}{c}{\textit{Interaction-Small}}&\multicolumn{2}{c}{\textit{Normal-Large}}&\multicolumn{2}{c}{\textit{Interaction-Large}} \\
			\hline \\[-1.8ex]
			& & Median & Mean & Median & Mean & Median & Mean & Median & Mean \\
			\hline \\[-1.8ex]
			\multicolumn{9}{l}{RB[\%]}\\
			\hline \\[-1.8ex]
			&CEP & $0.5737$ & $0.6575$ & $1.3044$ & $1.1705$ & $5.0801$ & $5.7415$ & $1.2549$ & $1.2620$ \\
			&GMERF & $0.2685$ & $0.2939$ & $0.0683$ & $0.0777$ & $5.1449$ & $5.6290$ & $1.5751$ & $1.6645$ \\
			&MQ & $2.1991$ & $2.2473$ & $2.4214$ & $2.3913$ & $7.7560$ & $7.9638$ & $4.5263$ & $4.6023$ \\
		 	\hline \\[-1.8ex]
		 	\multicolumn{9}{l}{RRMSE[\%]}\\
		 	\hline \\[-1.8ex]			
        &CEP & $10.3583$ & $10.4083$ & $15.6118$ & $15.7275$ & $27.3135$ & $29.9692$ & $23.7271$ & $23.7545$ \\
		&GMERF & $10.0811$ & $10.1047$ & $10.4459$ & $10.5813$ & $31.1783$ & $33.1685$ & $18.2625$ & $18.3164$ \\
		&MQ & $10.0061$ & $10.9667$ & $19.4710$ & $21.4834$ & $33.4434$ & $35.8279$ & $21.9926$ & $24.6999$ \\
		\hline \\[-1.8ex]
	\end{tabular}
\label{tab:MBpoint}
\end{table}

\begin{figure}[ht]
	\centering
	\captionsetup{justification=centering,margin=1.5cm}
	\includegraphics[width=\linewidth]{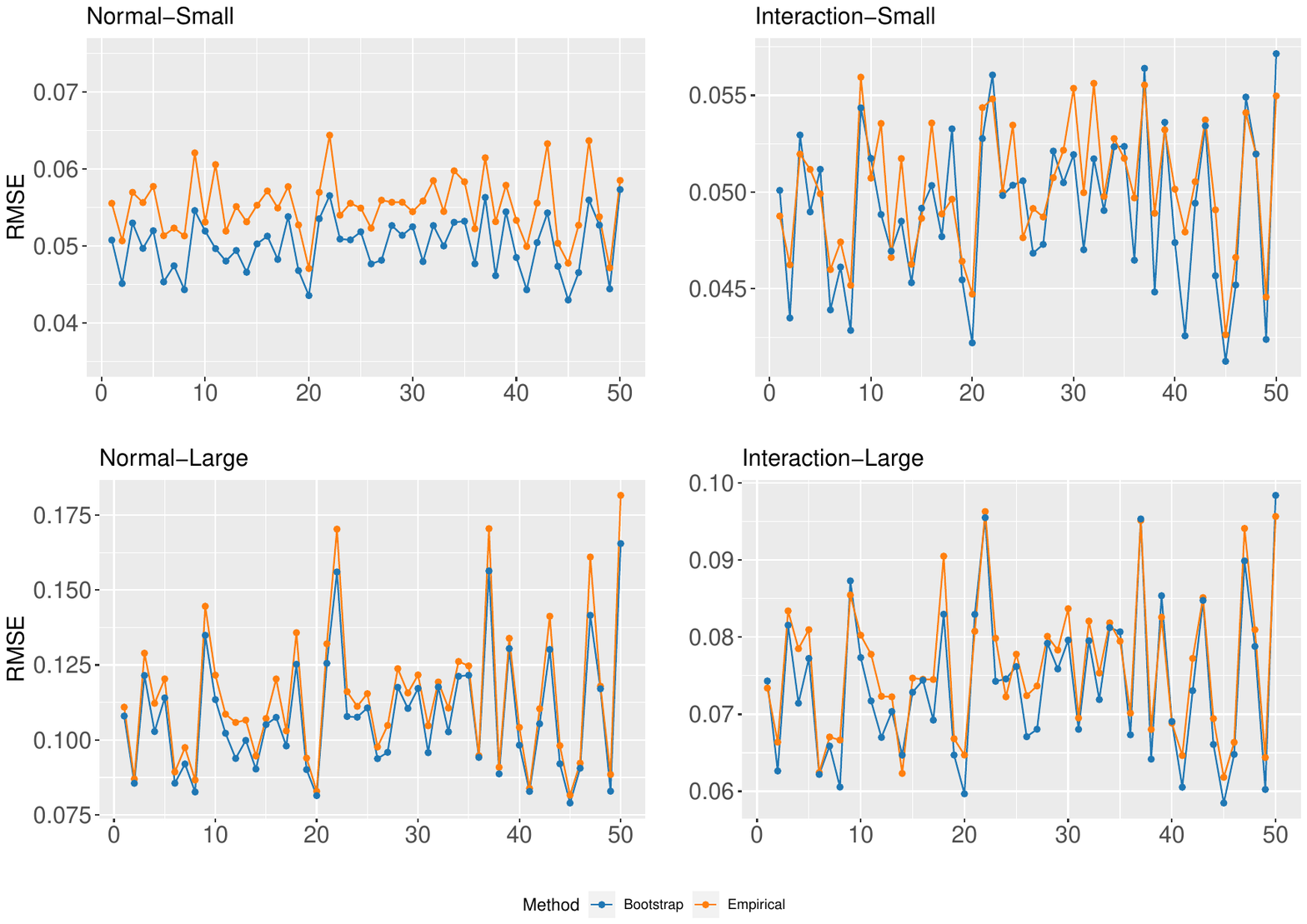}
	\caption{Empirical and bootstrapped area-level RMSEs for four scenarios}
	\label{fig:trackMSE}
\end{figure}

We now assess the performance of the MSE estimator implemented with the parametric bootstrap method presented in Section 3 with $B=200$ bootstrap replications. In Table 3, we observe the RB-RMSE of the proposed parametric bootstrap procedure across the four scenarios. In particular, the proposed MSE estimator demonstrates reasonably low relative bias in terms of mean and median values over areas under all four scenarios. Although we cannot directly infer the area-wise tracking properties of the estimated RMSE against the empirical RMSE from the results of Table \ref{tab:MBmse}, Figure \ref{fig:trackMSE} provides additional insight into the quality of our proposed parametric MSE-bootstrap estimator. Based on the tracking properties in all four scenarios, we conclude that using the parametric bootstrap for estimating the MSE appears to have appealing properties regarding bias and stability. 

\begin{table}[!h]
	\footnotesize
	\centering
	\captionsetup{justification=centering,margin=1.5cm}
	\caption{Performance of bootstrap MSE estimators in model-based simulation: mean and median of RB-RMSE and RRMSE-RMSE over areas}
	\begin{tabular}{@{\extracolsep{5pt}} lrcccccccc}
		\\[-1.8ex]\hline
		\hline \\[-1.8ex]
		& &\multicolumn{2}{c}{\textit{Normal-Small}} &\multicolumn{2}{c}{\textit{Interaction-Small}}&\multicolumn{2}{c}{\textit{Normal-Large}}&\multicolumn{2}{c}{\textit{Interaction-Large}} \\
		\hline \\[-1.8ex]
		& & Median & Mean & Median & Mean & Median & Mean & Median & Mean \\
		\hline \\[-1.8ex]
		&RB-RMSE[\%] & $-9.17$ & $-9.23$ & $-2.78$ & $-2.28$ & $-4.91$ & $-5.20$ & $-3.11$ & $-3.08$ \\
		 \\[-1.8ex]		
		&RRMSE-RMSE[\%] & $25.36$ & $25.72$ & $24.84$ & $25.40$ & $8.23$ & $8.75$ & $24.73$ & $24.98$ \\
		\hline \\[-1.8ex]
	\end{tabular}
	\label{tab:MBmse}
\end{table}

\section{Application: Estimating poverty rates for Mexican municipalities}\label{sec:5}
In this section, we investigate the performance implications of estimating the HCR using a binary variable versus a continuous income scale. Concretely, we compare the performance of the GMERF method with established methods for estimating poverty indicators in the context of SAE that use continuous household income as an input variable: the EBP with Log-Shift transformation \citep{Rojas2020} and the MERF for non-linear indicators \citep{Krennmair2022}. We apply these methods to Mexican income data to estimate area-specific HCRs and their associated uncertainties. Section 5.1 describes the data, while Section 5.2 presents the results.

\subsection{Data description}\label{sec:5.1}
Mexico has been plagued by poverty, affecting millions of children, men, women, the elderly, and the indigenous population in particular \citep{Conveal2020}. The country faces various obstacles in its development, including crime, governmental opacity, a scarcity of skilled labor, and corruption. Nearly half of the population lives below the poverty line, making poverty reduction a crucial priority. Since 2008, Mexico has implemented a multidimensional poverty measurement system, providing a more accurate assessment of social development policies at the federal, state, and municipal levels \citep{OECD2012}. However, the country has achieved mixed results in terms of reducing poverty rates. The social rights dimension has seen progress in terms of basic service coverage, such as education, health, housing, and social security. Meanwhile, the economic well-being dimension, as captured by individuals' wages, has experienced fluctuations over time, driven by significant events such as the financial crisis of 2008-2009. Subsequently, there was a period of recovery observed between 2014 and 2018. Nevertheless, the onset of the COVID-19 pandemic led to a subsequent decrease in economic well-being indicators \citep{Statista2021}. The picture is heterogeneous across the federal states, and this poses differentiated challenges to policymakers. Efforts have been made to refine the conceptual and methodological framework for measuring poverty, understand the living conditions of the population in poverty, and precisely identify their geographical location. As part of the Mexican State's institutional efforts to overcome poverty, the National Institute of Statistics and Geography (INEGI) designated the National Survey of Household Income and Expenditure (ENIGH) and the Socioeconomic Conditions Module of the ENIGH (MCS-ENIGH) as information of national interest. These data sources provide the necessary sample and auxiliary data required for estimating HCRs using SAE techniques. The ENIGH from 2010 serves as the sample data, while the census of 2010 provides the auxiliary data for the analysis. This study examines the regional differences in HCRs within one of the 32 Mexican federal states, Tlaxcala. Tlaxcala consists of 60 municipalities and had a population of 1342977 in 2020. It is the fifth least populated federal state, covering an area of 3996.6 $\text{km}^2$ and having a population density of 336 inhabitants per $\text{km}^2$ \citep{INEGI2021}. Additional geographic information (including the names of the municipalities) is represented in Figure A.1 and Table A.1.

The target variable for the EBP Log-Shift and MERF models is the total household per capita income (\textit{ictpc}), measured in pesos, which is available in the survey but not in the census. For the GMERF model, a binary target variable indicating whether a household is considered poor is created using the poverty line $t=0.60 \times median(\textit{ictpc})$, where households with $\textit{ictpc}_{ij} \leq t$ are classified as poor.

The census data for Tlaxcala from $2010$ contains information on $57751$ households and the survey data contains information on $1667$ households. Of the 60 municipalities, 52 are used in-sample and 8 are out-of-sample. Details on survey and census data properties are provided in Table \ref{tab:Apdetails}.  Regarding the survey, the maximum sample size of a municipality is $143$, the minimum is $2$ and the median is $21.50$ households per municipality. For the EBP Log-Shift model, we adopt the approach of \citet{Rojas2020} and use the Bayesian Information Criterion (BIC) to identify valid predictors for the target variable \textit{ictpc}. The GMERF algorithm's convergence is monitored under a precision of $1e^{-5}$ in the relative difference of the GLL criterion and with a precision of $0.01$ in the relative change in $\hat{\eta}$. The default of 500 trees is retained, and based on 5-fold cross-validation on the original survey-sample, it is recommended to use $2$ variables at each split for the GMERF.

\begin{table}[ht]
	\centering
	\captionsetup{justification=centering,margin=1.5cm}
	\caption{Summary statistics on in- and out-of-sample areas: area-specific sample size of census and survey data}
	\begin{tabular}{@{\extracolsep{5pt}} lcccccccc}
		\\[-1.8ex]\hline
		\hline \\[-1.8ex]
		&\multicolumn{2}{c}{Total}&\multicolumn{2}{c}{In-sample}&\multicolumn{2}{c}{Out-of-sample}\\
		&\multicolumn{2}{c}{60} & \multicolumn{2}{c}{52} & \multicolumn{2}{c}{8} \\ \hline
		\hline \\[-1.8ex]
		& Min. & Q1 & Median & Mean & Q3 & Max. \\
		\hline
		Survey area sizes & 2.00 & 11.00 & 21.50 & 32.06 & 40.25 & 143.00 \\
		Census area sizes & 611.00 & 772.20 & 912.00 & 962.59 & 1100.20 & 3310.00 \\
		\hline
	\end{tabular}
\label{tab:Apdetails}
\end{table}
In Section \ref{sec:5.2} and the subsequent design-based simulation in Section \ref{sec:5.3}, we present an illustrative and realistic example of the estimation of area-level HCRs using data from Tlaxcala. Our choice of this challenging example is deliberate, as it serves to demonstrate the validity of our proposed approaches for point and uncertainty estimates. Specifically, we aim to show that our method provides a viable alternative to existing SAE methods, while requiring less information about the input variables.

\subsection{Results and discussion}	\label{sec:5.2}
\begin{figure}[!htb]
	\centering
	\captionsetup{justification=centering,margin=1.5cm}
	\includegraphics[width=1\linewidth]{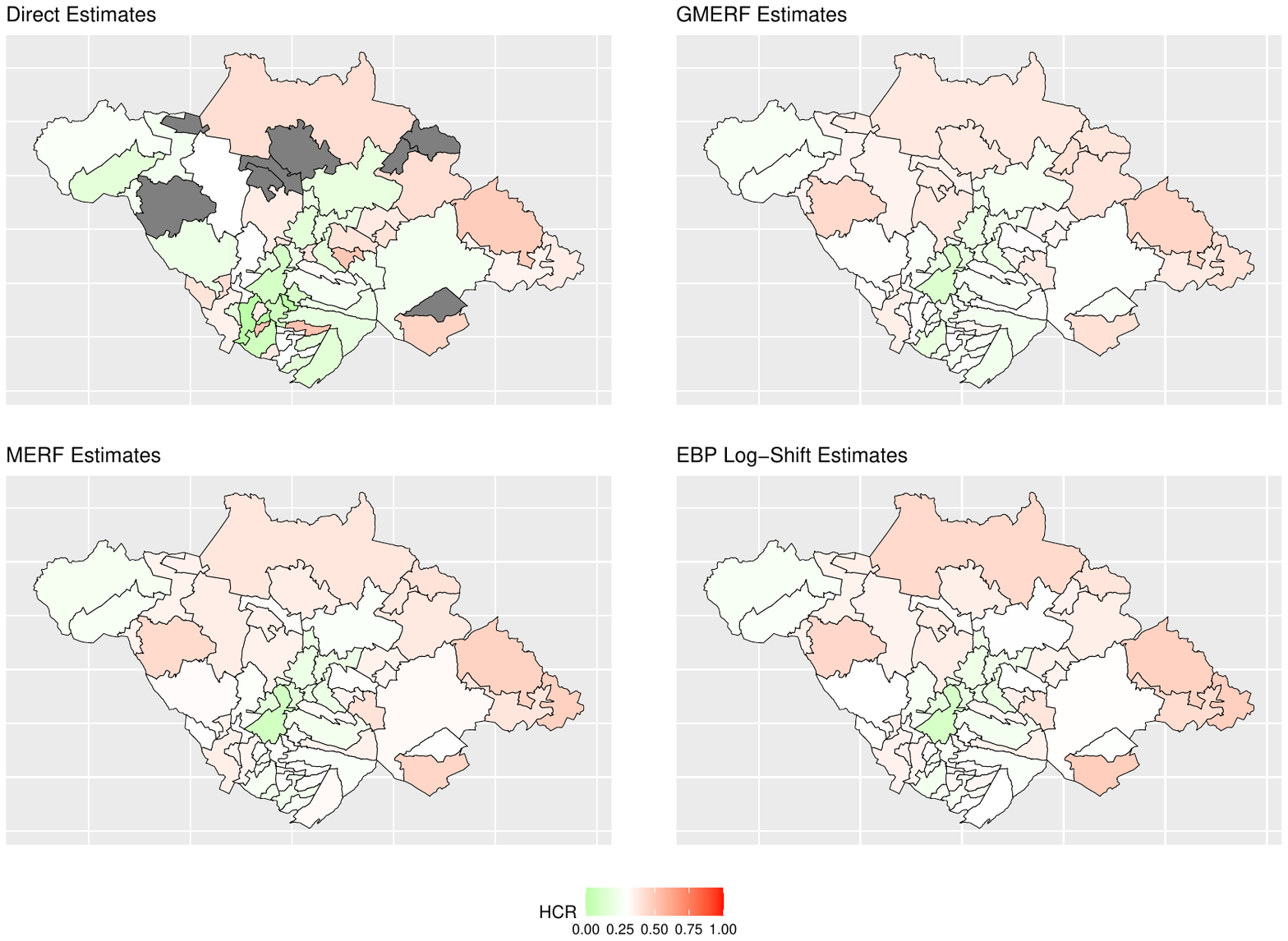}
	\caption{Estimated HCRs for the state Tlaxcala based on four different estimation methods}
	\label{fig:mapincome}
\end{figure}

Figure \ref{fig:mapincome} displays the results of direct estimates, as well as those from the model-based estimation methods (EBP Log-Shift, MERF and GMERF). Direct estimation of the HCRs is possible for 52 out of 60 domains. Clearly, the model-based estimates expand our understanding of regional disparities in HCRs beyond the sampled areas. All three model-based methods reveal distinct regional differences among municipalities. The state capital, Tlaxcala, located in the central Mexican highlands, has the lowest poverty rate. Income from tourism, given the state's rich history, could be one explanation for this phenomenon. Municipalities with slightly higher HCRs tend to be those with small commercial activities, such as Chiautempan, Humantla, San Pablo del Monte, and Zacatelco. Those with similarly high HCRs tend to be those with light industry, such as Atlangatepec, Calpulalpan, Ixtacuixtla de Mariano Matamoros, Nanacamilpa de Mariano Arista, Xicohtzinco, and Santa Isabel Xiloxoxtla, producing clothing, foam and plastic products, paper products, publishing, textiles, and automobiles. Municipalities in the north and east of the state tend to have the highest poverty rates, likely due to their rural nature and reliance on agriculture, livestock, forestry, and fishing \citep{INEGI2017}. Overall, the three model-based estimates consistently depict the geographic distribution of HCRs, regardless of whether they are based on a binary variable (GMERF) or a continuous variable (MERF and EBP Log-Shift).

\begin{figure}[!htb]	
	\centering
	\captionsetup{justification=centering,margin=1.5cm}
	\includegraphics[width=0.95\linewidth]{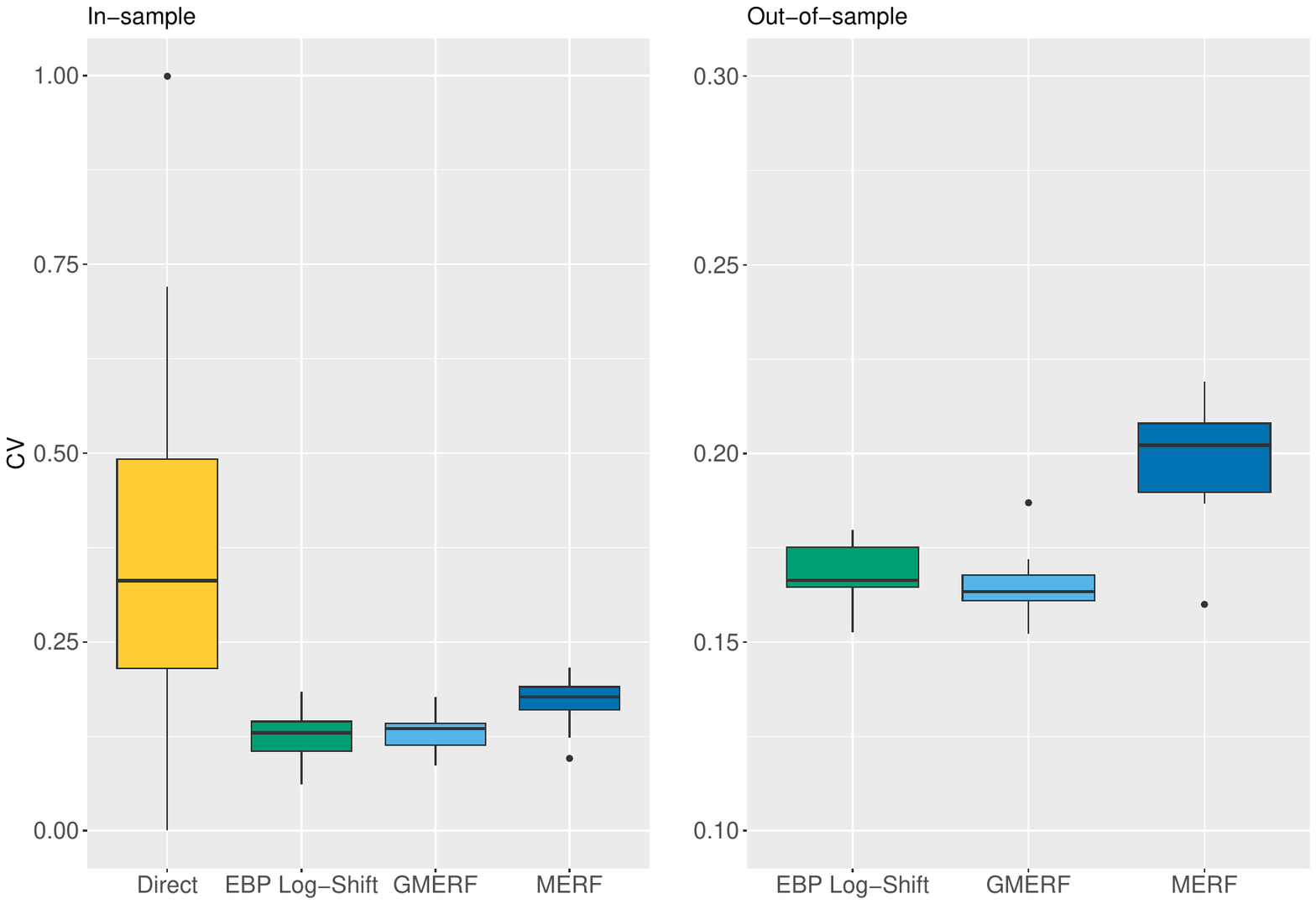}
	\caption{Domain-specific CVs for HCRs for in- and out-of-sample domains}
	\label{fig:detailCV}
\end{figure}

In addition to mapping the empirical results of domain proportions, our focus is on quality criteria, particularly the coefficients of variation (CV). We use the calibrated bootstrap method provided in the R package \emph{emdi} \citep{Kreutzmann_etal2019} to obtain estimates of variances for the direct estimates. To estimate the MSE for the EBP Log-Shift and MERF, we employ the wild-bootstrap and non-parametric bootstrap, respectively, as proposed by \citet{Rojas2020} and \citet{Krennmair2023}. For the GMERF, we rely on the parametric bootstrap from Section \ref{sec:3}, with $B=200$ bootstrap replications. The corresponding CVs for in- and out-of-sample domains are reported in Figure \ref{fig:detailCV}. We note an improvement in the in-sample CVs for the EBP Log-Shift, GMERF, and MERF compared to the CVs for the direct estimates. The median CVs for the EBP Log-Shift and GMERF are lower than those for the MERF. In terms of CVs for out-of-sample areas, we found that our proposed GMERF approach has an advantage. However, upon analyzing individual CV values, it is unclear whether the improved performance of GMERFs is due to superior point estimates for domain-level proportions or its relatively lower MSE-estimates.

\begin{figure}[!htb]	
	\centering
	\captionsetup{justification=centering,margin=1.5cm}
	\includegraphics[width=0.95\linewidth]{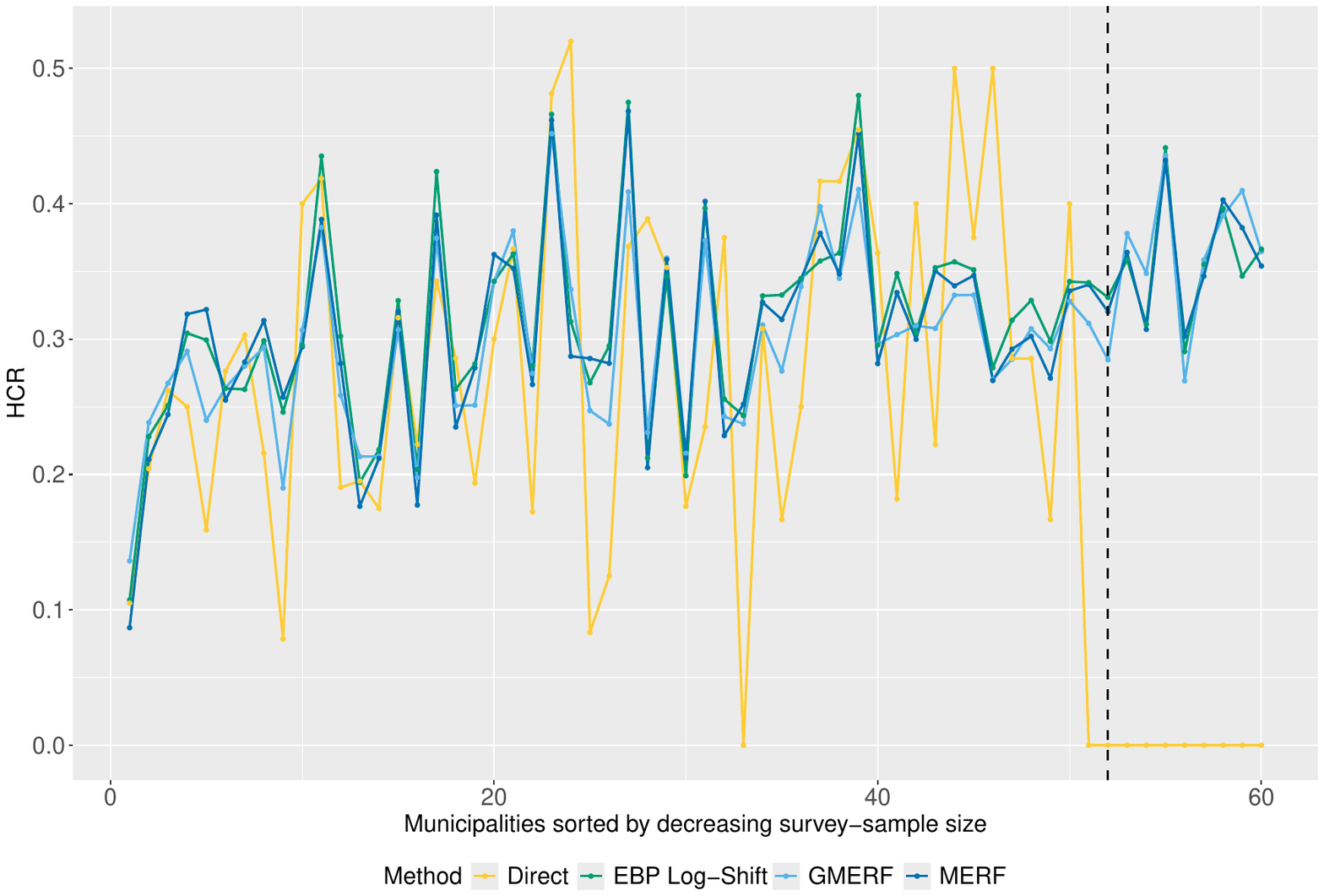}
	\caption{Detailed comparison of point estimates for the area-level HCRs. The dotted line separates sampled from non-sampled areas. In-sample areas are sorted by decreasing sample size}
	\label{fig:pointEstim}
\end{figure}

Thus, Figure \ref{fig:pointEstim} compares point estimates of direct estimates to model-based estimates for both in-sample and out-of-sample domains. Notably, there are no discernible systematic differences between the estimates from the EBP Log-Shift, GMERF, and the MERF. The in-sample areas in Figure \ref{fig:pointEstim} are sorted by decreasing sample sizes. In comparison to the direct estimates, the predicted HCRs of EBP Log-Shift, GMERF, and MERF exhibit less variation due to the impact of shrinkage.

\begin{figure}[ht]
	\centering
	\captionsetup{justification=centering,margin=1.5cm}
	\includegraphics[width=0.95\linewidth]{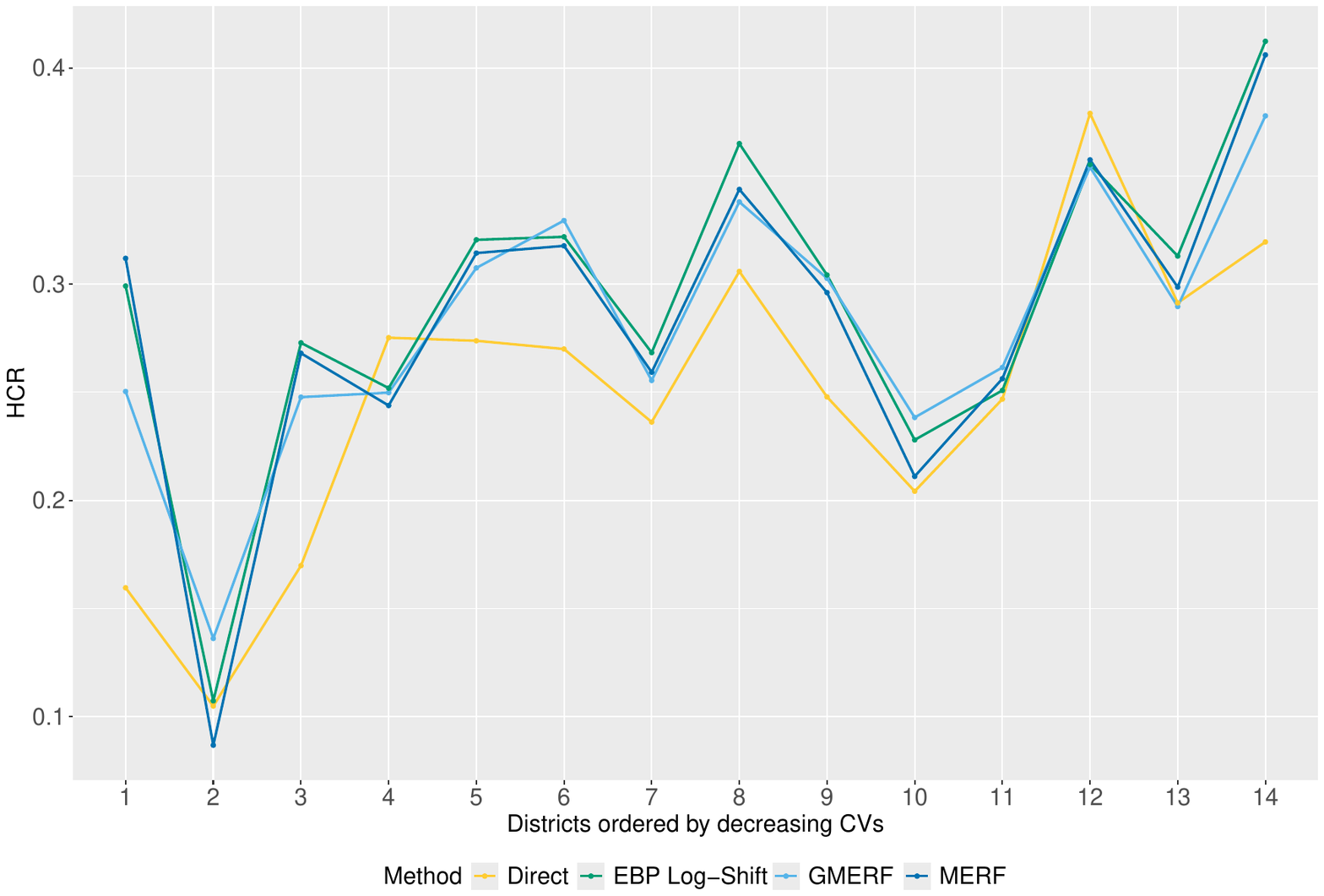}
	\caption{Estimates for HCRs at district level}
	\label{fig:pointEstimDis}
\end{figure}

Finally, we conduct an informal evaluation following the approach of \citet{Tzavidis2018} to compare model-based and design-based point estimates for aggregated geographical levels, providing an indication of the quality of the model-based estimates. Tlaxcala consists of 60 municipalities and 15 districts. The sample sizes range from 11 to 169 households per district, with a mean of 111 and a median of 109. We compare model-based estimates with design-based estimates for 14 districts for which the CVs of the design-based estimates are below 30\%. Figure \ref{fig:pointEstimDis} displays point estimates for district-level HCRs obtained using the direct estimator, EBP Log-Shift, MERF, and GMERF. Direct estimates are based on district-specific samples, while model-based estimates are aggregated from the corresponding municipality-level estimates using weights defined by $N_i/N$, where $N_i$ denotes the municipality population size. The districts are ordered by the CVs of the direct estimates from left to right. We observe that, for districts where the direct estimates are less reliable (left-hand side of the plot), the model-based estimates diverge more from the direct estimates. In contrast, for districts where the design-based estimates are more reliable (right-hand side of the plot), the GMERF estimates tend to be closer to the direct estimates. The correlation coefficient between the direct estimator and the GMERF ($0.87$) on district-level is higher than the correlation between the direct estimator and the EBP Log-Shift ($0.78$) and MERF ($0.76$). A design-based simulation will further validate our results and facilitate a more detailed discussion of our method.

\section{Design-based simulation study}\label{sec:5.3}
Regarding the design-based simulation, we are fortunate to have a variable in the survey and census datasets that is highly correlated with \textit{ictpc} in our application: the variable \textit{inglabpc}, which measures earned per capita income from work. Although this variable only covers one aspect of total household income and deviates from our desired income definition, it is effective in evaluating our method within a design-based simulation. By directly comparing the performance of the proposed GMERF approach to existing SAE methods for the estimation of area-level HCRs based on empirical data, the design-based simulation evaluates the results from the application.

We focus on area-level HCRs in the Mexican state of Tlaxcala and use the same data as in Section 5.2. We draw 500 independent pseudo-samples from the fixed population, maintaining the original survey's municipality sample sizes. This results in $500$ equally structured samples, each with an overall size of $1667$ households, as in the ENIGH 2010 survey. The true values are defined as the area-level HCRs from the  fixed population/ census. We employ the same methods as described in the application in Section \ref{sec:5.2} and use a consistent working model for EBP Log-Shift, assuming it remains fixed throughout the design-based simulation. Additionally, for EBP Log-Shift, GMERF, and MERF, we retain the parameters discussed in Section \ref{sec:5.2}.

\begin{figure}[ht]
	\centering
	\captionsetup{justification=centering,margin=1.5cm}
	\includegraphics[width=1\linewidth]{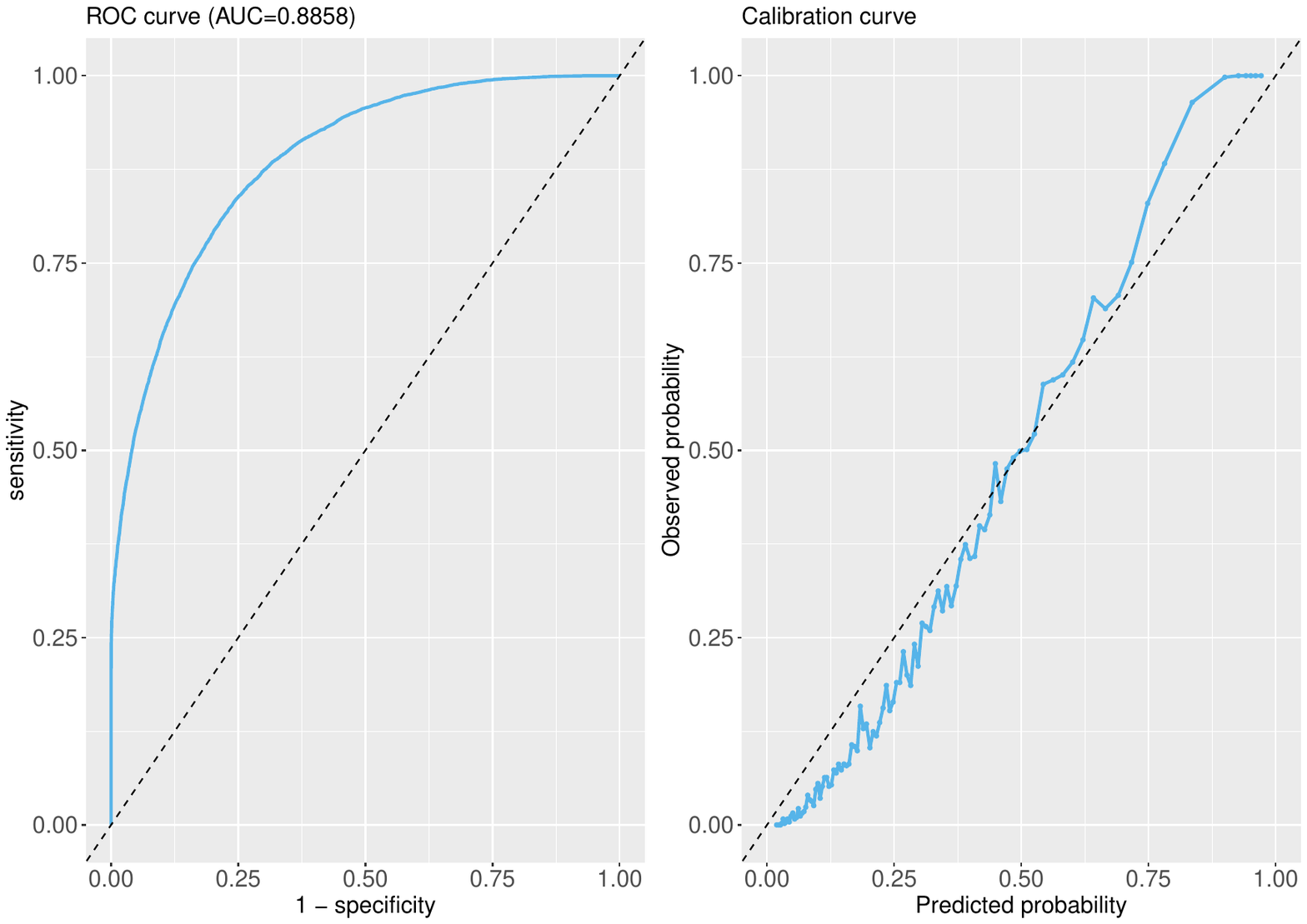}
	\caption{ROC and calibration curves of the GMERF}
	\label{fig:DBRoc}
\end{figure}

To start with, we examine the effectiveness of our method in correctly classifying households into the 'true' category (poor/rich) using the receiver operating characteristic (ROC) and calibration curves presented in Figure \ref{fig:DBRoc}. The ROC curve illustrates the relationship between the true positive rate (sensitivity) and the false positive rate ($1-$specificity), while the area under the curve (AUC) indicates the model's ability to differentiate between the two classes. A value closer to 1 indicates better class separation and prediction performance. In our case, we achieved an AUC of 0.8858, suggesting high classification and prediction accuracy. Additionally, the calibration curve on the right-hand side of Figure \ref{fig:DBRoc} displays the model's ability to accurately predict probabilities, by comparing the mean of predicted probabilities against the corresponding true probabilities at different quantiles. A diagonal line indicates a perfect match between the predicted and true probabilities. Our method demonstrates excellent performance in predicting probabilities within the $0.25$ to $0.75$ range, consistent with our theoretical considerations in Section 2.2.

\begin{figure}[ht]
	\centering
	\captionsetup{justification=centering,margin=1.5cm}
	\includegraphics[width=1\linewidth]{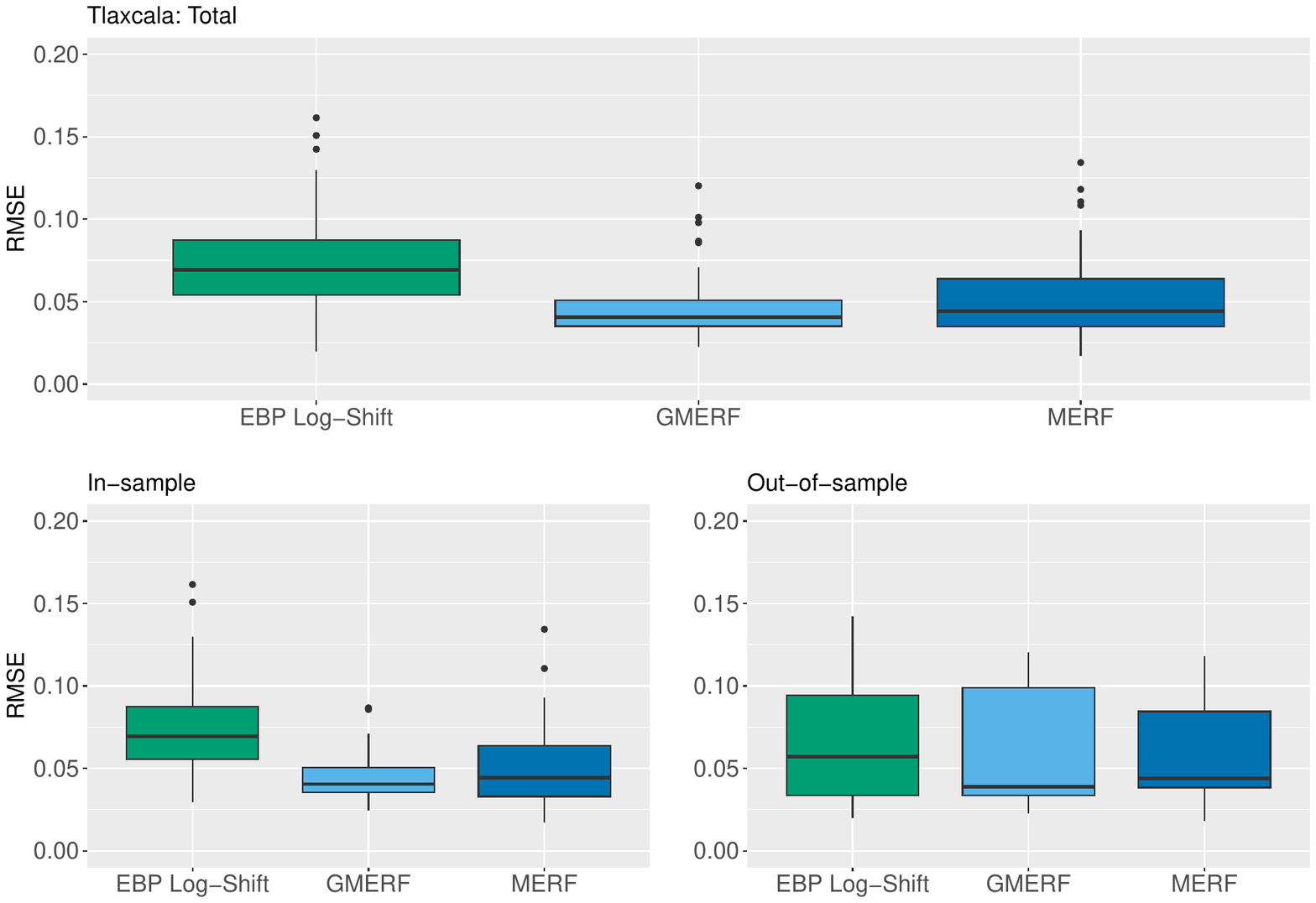}
	\caption{Performance of area-specific point estimates including details on in- and out-of-sample areas}
	\label{fig:DBpoint}
\end{figure}

We begin by discussing the performance of our method in terms of point estimation, comparing the EBP Log-Shift, GMERF, and MERF. Figure \ref{fig:DBpoint} displays the average RMSE of the area-level HCRs for Tlaxcala, both overall and split by the 52 in-sample and 8 out-of-sample areas. The GMERF method produces the lowest median RMSEs. However, for out-of-sample areas, there is only a small difference in the RMSEs of GMERF and MERF. Given the high share of sampled municipalities in Tlaxcala, the in-sample areas' performance is a critical indicator of each method's quality and stability. In this case, the non-parametric approaches of GMERF and MERF outperform the EBP Log-Shift in terms of RMSE. A direct comparison between GMERF and MERF shows that GMERF has better performance than MERF. These results are further supported by the discussion of mean and median values of bias and RRMSE in Table \ref{tab:DBpoint}. GMERF displays the lowest bias for all 60 areas, while for the 52 in-sample areas, GMERF outperforms its competitors in terms of both mean and median bias.
\begin{table}[!h]
	\centering
	\captionsetup{justification=centering,margin=1.5cm}
	\caption{Mean and median of Bias and RRMSE over in- and out-of-sample areas for point estimates}
	\begin{tabular}{@{\extracolsep{5pt}} lrcccccccc}
		\\[-1.8ex]\hline
		\hline \\[-1.8ex]
		& &\multicolumn{2}{c}{Total} &\multicolumn{2}{c}{In-sample}&\multicolumn{2}{c}{Out-of-sample} \\
		\hline \\[-1.8ex]
		& & Median & Mean & Median & Mean & Median & Mean \\
		\hline \\[-1.8ex]
		\multicolumn{7}{l}{Bias}\\
		\hline \\[-1.8ex]
		&EBP Log-Shift & $0.0639$ & $0.0611$  & $0.0661$ & $0.0695$& $0.0112$ & $0.0068$ \\
		&GMERF & $0.0282$ & $0.0188$ & $0.0288$ & $0.0239$ & $-0.0065$ & $-0.0148$ \\
		&MERF & $0.0343$ & $0.0283$ & $0.0370$ & $0.0339$& $-0.0158$ & $-0.0082$  \\
		\hline \\[-1.8ex]
		\multicolumn{7}{l}{RRMSE[\%]}\\
		\hline \\[-1.8ex]
		&EBP Log-Shift & $28.3263$ & $28.9663$ & $30.5159$ & $30.4340$ & $14.9920$ & $19.4268$ \\
		&GMERF & $16.6674$ & $18.8713$ & $16.6674$ & $18.8498$& $10.4226$ & $13.7361$  \\
		&MERF & $21.6265$ & $22.9136$  & $22.5494$ & $23.3961$& $11.6537$ & $16.2025$ \\
		\hline \\[-1.8ex]
	\end{tabular}
\label{tab:DBpoint}
\end{table}

Lastly, we assess the performance of the proposed parametric MSE-bootstrap procedure. Table \ref{tab:DBmse} presents the results of RB-RMSE and RRMSE-RMSE for the corresponding estimates. For in-sample areas, the median RB-RMSE indicates unbiasedness, while the mean RB-RMSE exhibits an acceptable level of overestimation. For out-of-sample areas, we observe a moderate level of overestimation leading to conservative MSE estimates. Overall, the proposed parametric bootstrap procedure meets the expectations, given the challenging conditions of this design-based simulation.
\begin{table}[!h]
	\centering
	\captionsetup{justification=centering,margin=1.5cm}
	\caption{Performance of MSE-estimator in design-based simulation: mean and median of RB and RRMSE over in- and out-of-sample areas}
	\begin{tabular}{@{\extracolsep{5pt}} lcccccccc}
		\\[-1.8ex]\hline
		\hline \\[-1.8ex]
		&\multicolumn{2}{c}{Total} &\multicolumn{2}{c}{In-sample}&\multicolumn{2}{c}{Out-of-sample} \\
		\hline \\[-1.8ex]
		& Median & Mean & Median & Mean & Median & Mean \\
		\hline \\[-1.8ex]
		RB-RMSE[\%] & $-0.1194$ & $8.0009$ & $-0.1242$ & $7.6599$ & $4.5545$ & $10.2179$ \\
		RRMSE-RMSE[\%] & $33.6429$ & $41.5158$ & $32.2307$ & $39.8012$ & $49.1337$ & $52.6610$ \\
		\hline \\[-1.8ex]
	\end{tabular}
\label{tab:DBmse}
\end{table}

\section{Conclusion}\label{sec:6}
In this paper, we propose GMERFs for the estimation of disaggregated binary-based poverty indicators. Additionally, we investigate the impact of information loss resulting from the conversion of continuous variables into binary variables on the performance of estimation methods. Our results suggest that even when confronted with limited information on income, our method has the potential to deliver comparable performance to methods that necessitate more detailed income data on a continuous scale.

The proposed GMERF model employs a combination of the PQL method with an algorithm reminiscent of the EM algorithm, allowing for a flexible specification of the model. The resulting estimator for area-level proportions is complemented by a modified parametric bootstrap scheme similar to \citet{Gonzalez_etal2008}. The performance of the point- and MSE estimates of the proposed method is evaluated against traditional SAE-methods for binary variables in a model-based simulation study. Furthermore, the proposed method is compared to established SAE-methods that use continuous household income as an input variable in an application and a design-based simulation study for estimating the HCR using income data from the Mexican state Tlaxcala. The model-based simulation demonstrates that our proposed point estimates perform well in scenarios with a linear specification and outperform traditional methods in the presence of non-linear interactions between covariates. The design-based simulation confirms the adequacy of GMERFs for point estimation under realistic conditions. We conclude that our proposed MSE-bootstrap scheme is reliable based on its performance in the model-based simulation, the application, and the design-based simulation.

From an applied perspective, additional research is required to establish a comprehensive metric indicating the extent to which SAE practitioners can forgo continuous income data and solely rely on rich or poor categorization to attain comparable estimators of point and uncertainty levels. A promising direction for future studies would involve extending the proposed approach to encompass count data. This expansion would enable the flexible estimation of a multidimensional poverty index, allowing for a more comprehensive assessment of poverty across multiple dimensions. Lastly, we propose a parametric bootstrap MSE for quantifying the uncertainty of the small area estimates. Exploring the development of non-parametric bootstrap and analytical estimators akin to those proposed by \citet{Krennmair2023} presents further prospects for future research.

\section*{Acknowledgements}
The authors are grateful to CONEVAL for providing the data used in empirical work. The views set out in this paper are those of the authors and do not reflect the official opinion of CONEVAL. The numerical results are not official estimates and are only produced for illustrating the methods. Additionally, the authors would like to thank the HPC Service of ZEDAT, Freie Universität Berlin, for computing time.

\clearpage

\bibliographystyle{apacite}				
\bibliography{./GMERF_paper_refs_1}

\clearpage

\begin{appendices}
	
	\renewcommand{\thesection}{A}
	
	\section{}

	\renewcommand{\thefigure}{A.\arabic{figure}}  
	\setcounter{figure}{0}
	\renewcommand{\thetable}{A.\arabic{table}}  
	\setcounter{table}{0}
	\vspace{-0.5cm}  
	\begin{figure}[h!]
		\centering
		\captionsetup{justification=centering, margin=1.5cm}
		\includegraphics[width=0.7\textheight]{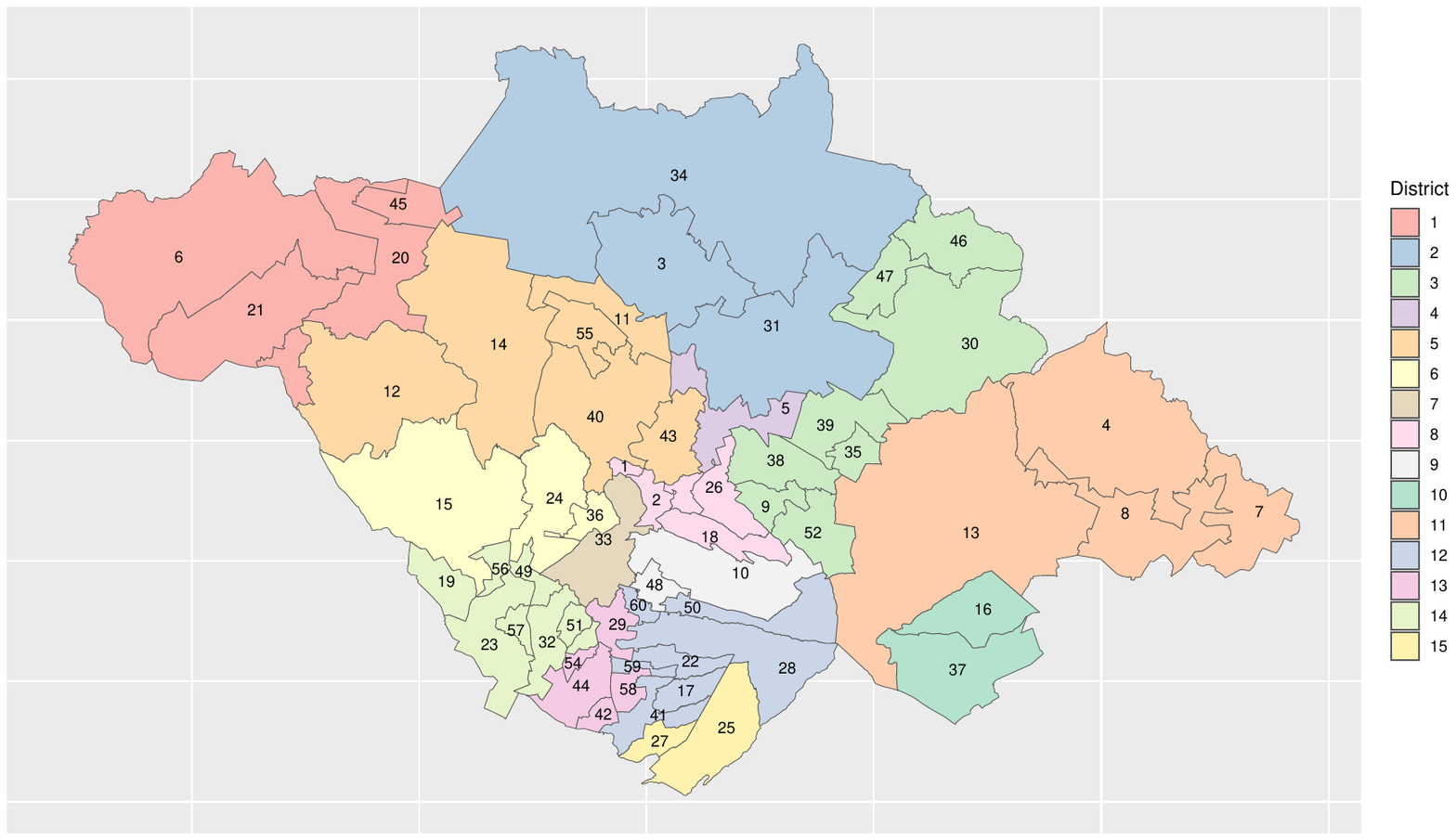}
		\caption{Municipalities and districts in Tlaxcala}
	\end{figure}
	
\begin{longtable}{cl}
	\caption{List of the municipalities in Tlaxcala} \\
	\hline \hline \hline
	ID & Municipality\\ \hline
	\endfirsthead
	
	\multicolumn{2}{c}%
	{\tablename\ \thetable\ -- \textit{Continued from previous page}} \\
	\hline \hline
	ID & Municipality\\ \hline
	\endhead
	
	\hline \hline
	\multicolumn{2}{r}{\textit{Continued on next page}} \\
	\endfoot
	
	\hline \hline
	\endlastfoot
	
	1 & Amaxac de Guerrero\\
	2 & Apetatitlán de Antonia Carvajal\\
	3 & Atlangatepec\\
	4 & Altzayanca\\
	5 & Apizaco\\
	6 & Calpulalpan\\
	7 & El Carmen Tequexquitla\\
	8 & Cuapiaxtla\\
	9 & Cuaxomulco\\
	10 & Chiautempan\\
	11 & Mu$\tilde{\text{n}}$oz de Domingo Arenas\\
	12 & Espa$\tilde{\text{n}}$ita\\
	13 & Huamantla\\
	14 & Hueyotlipan\\
	15 & Ixtacuixtla de Mariano Matamoros\\
	16 & Ixtenco\\
	17 & Mazatecochco de José Maria Morelos\\
	18 & Contla de Juan Cuamatzi\\
	19 & Tepetitla de Lardizábal\\
	20 & Sanctorum de Lázaro Cárdenas\\
	21 & Nanacamilpa de Mariano Arista\\
	22 & Acuamanala de Miguel Hidalgo\\
	23 & Nativitas\\
	24 & Panotla\\
	25 & San Pablo del Monte\\
	26 & Santa Cruz Tlaxcala\\
	27 & Tenancingo\\
	28 & Teolocholco\\
	29 & Tepeyanco\\
	30 & Terrenate\\
	31 & Tetla de la Solidaridad\\
	32 & Tetlatlahuca\\
	33 & Tlaxcala\\
	34 & Tlaxco\\
	35 & Tocatlan\\
	36 & Totolac\\
	37 & Zitlaltepec de Trinidad Sánchez Santos\\
	38 & Tzompantepec\\
	39 & Xaloztoc\\
	40 & Xaltocan\\
	41 & Papalotla de Xicohtencatl\\
	42 & Xicohtzinco\\
	43 & Yauhquemecan\\
	44 & Zacatelco\\
	45 & Benito Juárez\\
	46 & Emiliano Zapata\\
	47 & Lázaro Cárdenas\\
	48 & La Magdalena Tlaltelulco\\
	49 & San Damián Texoloc\\
	50 & San Francisco Tetlanohcan\\
	51 & San Jerónimo Zacualpan\\
	52 & San José Teacalco\\
	53 & San Juan Huactzinco\\
	54 & San Lorenzo Axocomanitla\\
	55 & San Lucas Tecopilco\\
	56 & Santa Ana Nopalucan\\
	57 & Santa Apolonia Teacalco\\
	58 & Santa Catarina Ayometla\\
	59 & Santa Cruz Quilehtla\\
	60 & Santa Isabel Xiloxoxtla\\
	\hline
\end{longtable}

\end{appendices}

\end{document}